\documentclass[%
 reprint,
nofootinbib,
 amsmath,amssymb,
 aps,
]{revtex4-1}

\usepackage{hyperref}
\usepackage{todonotes}
\usepackage{graphicx}
\usepackage{dcolumn}
\usepackage{bm}
\usepackage{chemformula}
\usepackage[utf8]{inputenc}

\hypersetup{colorlinks,linkcolor={blue},citecolor={teal},urlcolor={violet}}

\begin{document}

\title{A Unique Multi-Messenger Signal of QCD Axion Dark Matter}

\author{Thomas D. P. Edwards}
\email{t.d.p.edwards@uva.nl}
\author{Marco Chianese}
\email{m.chianese@uva.nl}
\author{Bradley J. Kavanagh}
\email{b.j.kavanagh@uva.nl}
\author{Samaya M. Nissanke}
\email{s.m.nissanke@uva.nl}
\author{Christoph Weniger}
\email{c.weniger@uva.nl}
\affiliation{%
 Gravitation Astroparticle Physics Amsterdam (GRAPPA), Institute
for Theoretical Physics Amsterdam and Delta Institute for Theoretical Physics,
University of Amsterdam, Science Park 904, 1090 GL Amsterdam, The Netherlands 
}%

\date{\today}

\begin{abstract}
We propose a multi-messenger probe of QCD axion Dark Matter based on observations of black hole-neutron star binary inspirals. It is suggested that a dense Dark Matter spike may grow around intermediate mass black holes ($10^{3}-10^{5} \mathrm{\,M_{\odot}}$). The presence of such a spike produces two unique effects: a distinct phase shift in the gravitational wave strain during the inspiral and an enhancement of the radio emission due to the resonant axion-photon conversion occurring in the neutron star magnetosphere throughout the inspiral and merger. Remarkably, the observation of the gravitational wave signal can be used to infer the Dark Matter density and, consequently, to predict the radio emission. We study the projected reach of the LISA interferometer and next-generation radio telescopes such as the Square Kilometre Array. Given a sufficiently nearby system, such observations will potentially allow for the detection of QCD axion Dark Matter in the mass range $10^{-7}\,\mathrm{eV}$ to $10^{-5}\,\mathrm{eV}$.

\end{abstract}

\pacs{Valid PACS appear here}
\maketitle


\textbf{\textit{Introduction ---}} The particle nature of Dark Matter (DM) remains a mystery~\cite{Bertone:2004pz,Bertone:2018xtm} despite efforts to detect it through astrophysical and lab-based observations~\cite{Duffy:2009ig,Undagoitia:2015gya,Gaskins:2016cha,Kahlhoefer:2017dnp}. Another indication of New Physics comes from the Strong charge parity (CP) problem of quantum chromodynamics (QCD) \cite{Dine:2000cj}. The non-observation of the neutron electric dipole moment~\cite{Afach:2015sja} constrains the CP-violating $\theta$-parameter in the QCD sector to be surprisingly small, $\left|\theta\right|\lesssim 10^{-10}$, while it could generically be $\mathcal{O}(1)$. A popular solution to this fine-tuning issue is the Peccei-Quinn mechanism, which predicts the existence of the {\it axion}~\cite{Peccei:1977hh,Peccei:1977ur,Weinberg:1977ma,Wilczek:1977pj}. Axion-like particles are also predicted in several extensions of the Standard Model, as well as in string theory~\cite{Arvanitaki:2009fg}. However, in the case of the QCD axion there a tight relation between its mass and its couplings with ordinary matter~\cite{Kim:1979if,Shifman:1979if,Zhitnitsky:1980tq,Dine:1981rt}.

These two fundamental issues can be addressed simultaneously by treating the QCD axion as DM~\cite{Preskill:1982cy,Abbott:1982af,Dine:1982ah} (see Ref.~\cite{Marsh:2015xka} for a review). Axions may be produced with the correct relic abundance through the misalignment mechanism~\cite{Wantz:2009it}, through the decay of topological defects such as strings and domain walls~\cite{Davis:1986xc, Harari:1987ht, Hagmann:2000ja, Hiramatsu:2012gg,Kawasaki:2014sqa}, or via thermal production~\cite{Turner:1986tb,Salvio:2013iaa}. So far, only a small part of the QCD axion parameter space has been explored 
~\cite{Asztalos:2009yp,Du:2018uak, Anastassopoulos:2017ftl,Hoof:2018ieb}, though new experimental
techniques have been recently proposed~\cite{Brubaker:2016ktl,TheMADMAXWorkingGroup:2016hpc,Zhong:2018rsr,Brun:2019lyf,Lawson:2019brd,Kahn:2016aff,McAllister:2017lkb,Shokair:2014rna,Kenany:2016tta,Alesini:2017ifp,Caputo:2018ljp,Caputo:2018vmy} (see Ref.~\cite{Irastorza:2018dyq} for a comprehensive review). Furthermore, it has been noted that the Primakov effect can efficiently convert axions to photons in the magnetic fields of Neutron Stars (NSs)~\cite{Pshirkov:2007st}. These photons are potentially observable with current and future radio telescopes
~\cite{Huang:2018lxq,Hook:2018iia,Safdi:2018oeu}.
\begin{figure}[t!]
    \centering
    \includegraphics[width=1.0\linewidth]{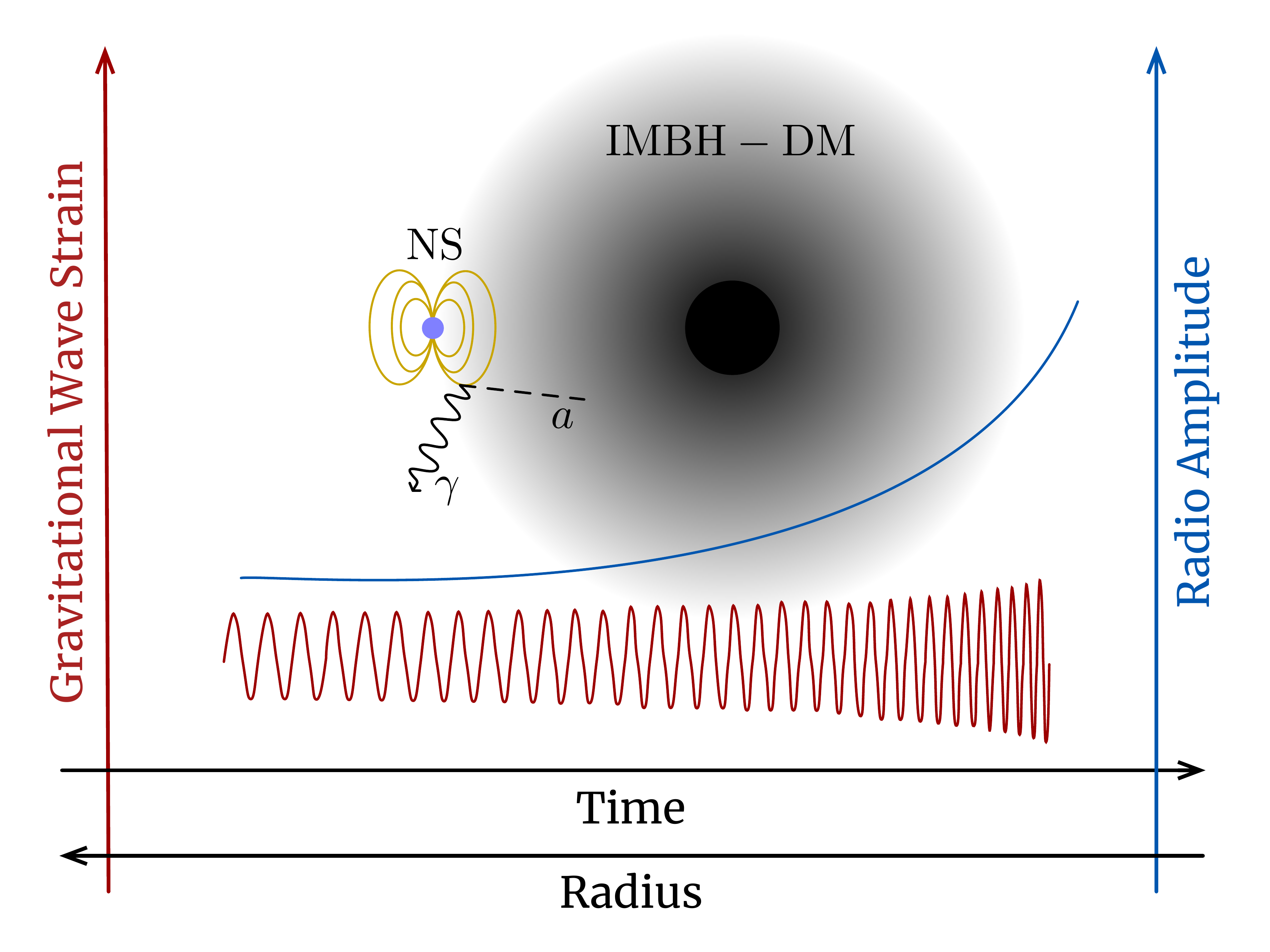}
    \caption{\textbf{Illustration of the IMBH-DM-NS system.}
    A DM halo of axions $a$ around the intermediate mass black hole (IMBH) produces a phase shift in the GW signal,
    and radio emission due to its conversion into photons $\gamma$ in the neutron star (NS) magnetosphere.}
    \label{fig:sketch}
\end{figure}

Gravitational Waves (GWs) has provided a new observational portal into extreme astrophysical environments \cite{Abbott:2016blz}. The detection of the binary NS merger GW170817 and 
electromagnetic counterparts further revolutionised astrophysics and multi-messenger astronomy~\cite{TheLIGOScientific:2017qsa,GBM:2017lvd}.
Distortions to GWs from binary black holes, caused by finite size effects of superradiant clouds, have recently been shown to provide a new probe of beyond the Standard Model (BSM) physics ~\cite{Baumann:2018vus,Hannuksela:2018izj}. References~\cite{Eda:2013gg,Eda:2014kra} demonstrated that 
a DM mini-spike around an intermediate mass black hole (IMBH) can dramatically affect the GW waveform through dynamical friction, providing yet another direct probe of BSM physics. DM environmental effects on GW signals were studied more generally in Refs.~\cite{Macedo:2013qea,Barausse:2014tra,Yue:2017iwc}.
 
In this \textit{Letter}, we explore the possibility of probing QCD axion DM with multi-messenger astronomy. We study the combined signal of GWs and radio emission from a NS inspiraling towards an IMBH surrounded by a dense spike of axion DM as sketched in Fig.~\ref{fig:sketch}. We show that by measuring the spike profile from the GW signal using the planned space-based observatory LISA~\cite{Audley:2017drz} (Fig.~\ref{fig:density}), we can predict the gradual evolution of the radio signal during the years-long inspiral phase. Most importantly, the increased density from the mini-spike amplifies the radio signal,
allowing for the potential detection of QCD axions with photon couplings expected in the most commonly studied models~\cite{Kim:1979if,Shifman:1979if,Zhitnitsky:1980tq,Dine:1981rt}. This is illustrated for 
radio observations with the Square Kilometre Array (SKA)~\cite{Bull:2018lat,SKA} 
in Fig.~\ref{fig:limit}.

\medskip
\textbf{\textit{Astrophysical System ---}}
IMBHs 
have masses $M_{\text{BH}} =10^3-10^5 M_{\odot}$. Thought to reside in the centres of smaller spiral galaxies ($\lesssim 10^9\, M_{\odot}$~\cite{2012NatCo...3.1304G}), as well as in dense stellar environments such as globular clusters \cite{Miller:2001ez}, a growing number of observations point toward the existence of IMBHs in nature \cite{Webb:2013sja,2018MNRAS.480.4684B,2019ApJ...871L...1T,2019arXiv190500145W}. There are multiple proposed formation mechanisms, including runaway growth through the mergers of stellar mass objects \cite{Taniguchi:2000mp,PortegiesZwart:2002iks,PortegiesZwart:2004ggg}; the direct collapse of gas clouds at high redshift \cite{Begelman:2006db,2012MNRAS.425.2854A}; or primordial formation from large density perturbations~\cite{Carr:1975qj,Kawaguchi:2007fz,Carr:2018poi,Carr:2019kxo}. 

These IMBHs may exist in DM halos~\cite{Islam:2002gg,Zhao:2005zr,Bertone:2005xz,Rashkov:2013uua}. It has been shown that for a BH undergoing adiabatic growth \cite{Sigurdsson:2003wu} at the centre of such a halo, the surrounding DM would form a dense spike whose profile $\rho_{\text{DM}}(r)$ is approximately a power law with index $\alpha$ \cite{Blumenthal:1985qy,Quinlan:1994ed,Gondolo:1999ef, Bertone:2002je,Bertone:2009kj,Sadeghian:2013laa}:
\begin{equation}
\rho_{\text{DM}}(r) =
     \begin{cases}
       \rho_{\text{sp}}\left(\frac{r_{\text{sp}}}{r}\right)^{\alpha}\,, &\quad r_{\text{ISCO}} < r \leq r_{\text{sp}}\\
       \frac{\rho_\text{s}}{(r/r_\text{s})(1+r/r_\text{s})^2}\,, &\quad \qquad r > r_{\text{sp}}\,.\\
     \end{cases}
\end{equation}
The NFW parameters $\rho_\text{s}$ and $r_\text{s}$ \cite{Navarro:1995iw} are related to the cosmological history and mass of the halo, for which we follow Ref.~\cite{Eda:2014kra}, assuming a formation redshift $z_f = 20$ and total halo mass $10^6\,M_\odot$. The radius of the BH's inner-most stable circular orbit (ISCO) is denoted $r_\mathrm{ISCO}$. To solve for the spike parameters we set $M(<r_{\mathrm{h}}) = 4\pi\int_{r_\mathrm{ISCO}}^{r_{\mathrm{h}}}\rho_{\mathrm{DM}}r^2\,\mathrm{d}r = 2M_{\mathrm{BH}}$, where  $r_{\mathrm{sp}} \sim 0.2 \, r_\mathrm{h}$~\cite{Eda:2014kra}. The spike profile varies with the initial DM profile. For an initially NFW-like profile, the spike slope is $\alpha=7/3$ (our benchmark scenario).

These systems are speculative from both the perspective of IMBH formation as well as the presence and properties (such as $\alpha$) of the spike~\cite{Ullio:2001fb,Merritt:2002vj,Zhao:2005zr,Hannuksela:2019vip}. For instance, for the spike to be preserved, the BH must not have undergone any mergers in its recent past~\cite{Merritt:2002vj}, nor should it be in a dense baryonic environment~\cite{Ullio:2001fb}. So if these systems form, their most likely location is in globular clusters~\cite{Fragione:2017blf}.

In addition to the IMBH with a DM-spike, we consider an inspiraling NS (on a circular orbit, for concreteness). NSs can have extremely high magnetic fields ($10^{9} - 10^{15} \mathrm{\,G}$), allowing for efficient axion-photon conversion close to the NS surface. NSs readily form in globular clusters \cite{2013IAUS..291..243F} and are therefore plausible candidates for mergers with IMBHs. We refer to the total system as IMBH-DM-NS.
 
Reference~\cite{Zhao:2005zr} argues that there are many IMBHs within our own Galactic halo. For an IMBH-DM-NS system to form, the IMBH must capture a NS. This process is very uncertain, relying on tracing formation models from the early Universe to today \cite{Mandel:2007hi,Fragione:2017blf}. Reference~\cite{Fragione:2017blf} suggests a detection rate density in LISA of approximately $\mathcal{R} \sim 3-10 \,\mathrm{Gpc^{-3}\,yr^{-1}}$. We therefore consider two scenarios, one in which  the IMBH-DM-NS system is close, at $0.01$~Gpc, and one in which the system is further away, at $1$~Gpc. The former is an optimistic scenario in terms of the strength of the radio signal, whereas many of the farther systems are likely to be observed over a ten year observing period.\footnote{Note that 1 Gpc corresponds to $z \approx 0.25$ and a signal-to-noise ratio of $\sim 1$ in LISA.} Importantly, these events would be dominated by IMBHs with masses $10^3-10^4\,\mathrm{M}_{\odot}$. For concreteness we consider an IMBH of $10^4\,\mathrm{M}_{\odot}$ since the additional gravitational potential of the BH preserves the structure of the spike for longer times \cite{Kavanagh:2020cfn}.

The properties of NSs in globular clusters are uncertain. However, they are thought to be much older than normal pulsars in galactic disks and it has been suggested that most are formed from electron-capture supernova processes due to their minimal kick velocities \cite{2013IAUS..291..243F,Safdi:2018oeu}. For the inspiraling NS, we take the magnetic field strength $B_0 = 10^{12} \mathrm{\,G}$ and spin period $P = 10 \mathrm{\,s}$~\cite{2013IAUS..291..243F,Hook:2018iia}; NSs with similar properties have been found in observed globular clusters \cite{1996ApJ...460L..41L,1994MNRAS.267..125B,1993Natur.361...47L,2008MNRAS.386..553I,2011ApJ...742...51B}. We assume $M_{\rm NS} = 1.4\,\mathrm{M}_{\odot}$ and $r_{\rm NS} = 10$~km as benchmark values for the NS mass and radius.

\medskip
\textbf{\textit{Gravitational Wave signal ---}} In vacuum, intermediate mass-ratio inspiral produces sub-Hertz gravitational waves. In the IMBH-DM-NS system, the dominant effect causing a deviation from the vacuum inspiral is the gravitational interaction between the DM halo and the NS passing through it, known as dynamical friction (DF)~\cite{Chandrasekhar1943a,Chandrasekhar1943b,Chandrasekhar1943c}. Dynamical friction exerts a drag force on the NS:
\begin{equation}
    f_{\text{DF}} = 4 \pi G_N^2 M_\mathrm{NS}^2 \frac{\rho_{\text{DM}}(r)}{v_{\text{NS}}^2(r)} \ln \Lambda\,,
\label{eq:DynamicalFriction}
\end{equation}
where $v_{\text{NS}}$ is the velocity of the NS, and  we take $\ln \Lambda \sim 3$ for the Coulomb logarithm. This force causes a loss of orbital energy, $\mathrm{d} E_{\mathrm{DF}}/\mathrm{d} t = v_\mathrm{NS} f_\mathrm{DF}$,
changing the accumulated phase of the GW signal and eventually reducing the inspiral time before merger with respect to the vacuum waveform. We see from Eq.~\eqref{eq:DynamicalFriction} that this force grows as the NS inspirals,\footnote{The NS orbital velocity grows roughly as $r^{-1/2}$, so that the DF force scales roughly as $r^{-\alpha + 1}$.} although so too does the radiation reaction force due to GW emission.

In the Newtonian regime, the 
waveform of the IMBH-DM-NS system is computed by solving the energy balance equation, taking into account the effect of both DF and GW emission on the orbital energy $E_{\text {orbit }}$ of the system~\cite{Eda:2014kra}:
\begin{equation}
    -\frac{\mathrm{d} E_{\text {orbit }}}{\mathrm{d} t}=\frac{\mathrm{d} E_{\mathrm{DF}}}{{\rm d} t} + \frac{\mathrm{d} E_{\mathrm{GW}}}{\mathrm{d} t}\,.
\end{equation}
For circular orbits in the Newtonian regime, the energy loss due to GW emission is
\begin{equation}
\frac{\mathrm{d} E_{\mathrm{GW}}}{\mathrm {d} t}=\frac{32}{5} \frac{G_N\, M_{\mathrm{NS}}^{2}}{c^{5}} r^{4} \, \omega_{s}^{6}\,,
\end{equation}
where $\omega_{s}$ is the orbital frequency and $r$ is the orbital radius. The resulting phase difference with respect to the vacuum inspiral signal depends on the chirp mass $\mathcal { M }_c \simeq M _ \mathrm{ NS }^ { 3 / 5 }  \,M _ \mathrm{BH } ^ { 2 / 5 } $, on the individual masses $M_\mathrm{BH}$ and $M_\mathrm{NS}$, and on the density of DM.\footnote{Note that we only consider low redshift systems, $z \ll 1$, therefore we ignore any difference between lab and system frame.}

Figure~\ref{fig:density} shows the constraints on $\alpha$  as a function of radius from the IMBH recast as an error on the DM density. To calculate the error we take ten log-spaced radial bins and integrate the noise-weighted inner product between the associated lower and upper frequencies, $f^i_l$ and $f^i_u$ respectively. The error on $\alpha$ \cite{Cutler:1994ys} (using the Fisher information) is then given by:
\begin{equation}
    \frac{\Delta\alpha}{\alpha} = \left[4\, \mathrm{Re}\left(\int^{f^i_u}_{f^i_l} \frac{\frac{\partial h}{\partial \ln \alpha} \frac{\partial h^{*}}{\partial \ln \alpha}}{S_n(f)} \,\mathrm{d}f\right) \right]^{-\frac{1}{2}} \,,
\end{equation}
where $S_n(f)$ is the LISA noise spectral density taken from Ref.~\cite{Eda:2014kra} and $h$ is the GW strain. We assume a 5-year observation with LISA, beginning at a frequency of 0.04 Hz at $r \approx 1.5 \times 10^{-8}\,\mathrm{pc}$ and ending at 0.44 Hz at the ISCO. 

We neglect any errors from the correlation between different parameters, which are expected to be small for $\alpha$~\cite{Eda:2014kra}. We assume that all quantities (for example spins and masses) can be measured precisely and do not contribute significantly to the error on the DM density. Note that higher order post-Newtonian effects on the inspiral will be important in breaking the degeneracy between $M_\mathrm{BH}$, $M_\mathrm{NS}$ and $\mathcal{M}_c$, as well as deducing the spins of the NS and IMBH. This degeneracy breaking has been demonstrated for current and future experiments in Refs.~\cite{TheLIGOScientific:2016pea,Barack:2003fp,TheLIGOScientific:2016wfe}, although the impact on our projections should be tested in future work.
\begin{figure}[t!]
    \centering
    \includegraphics[width=1.0\linewidth]{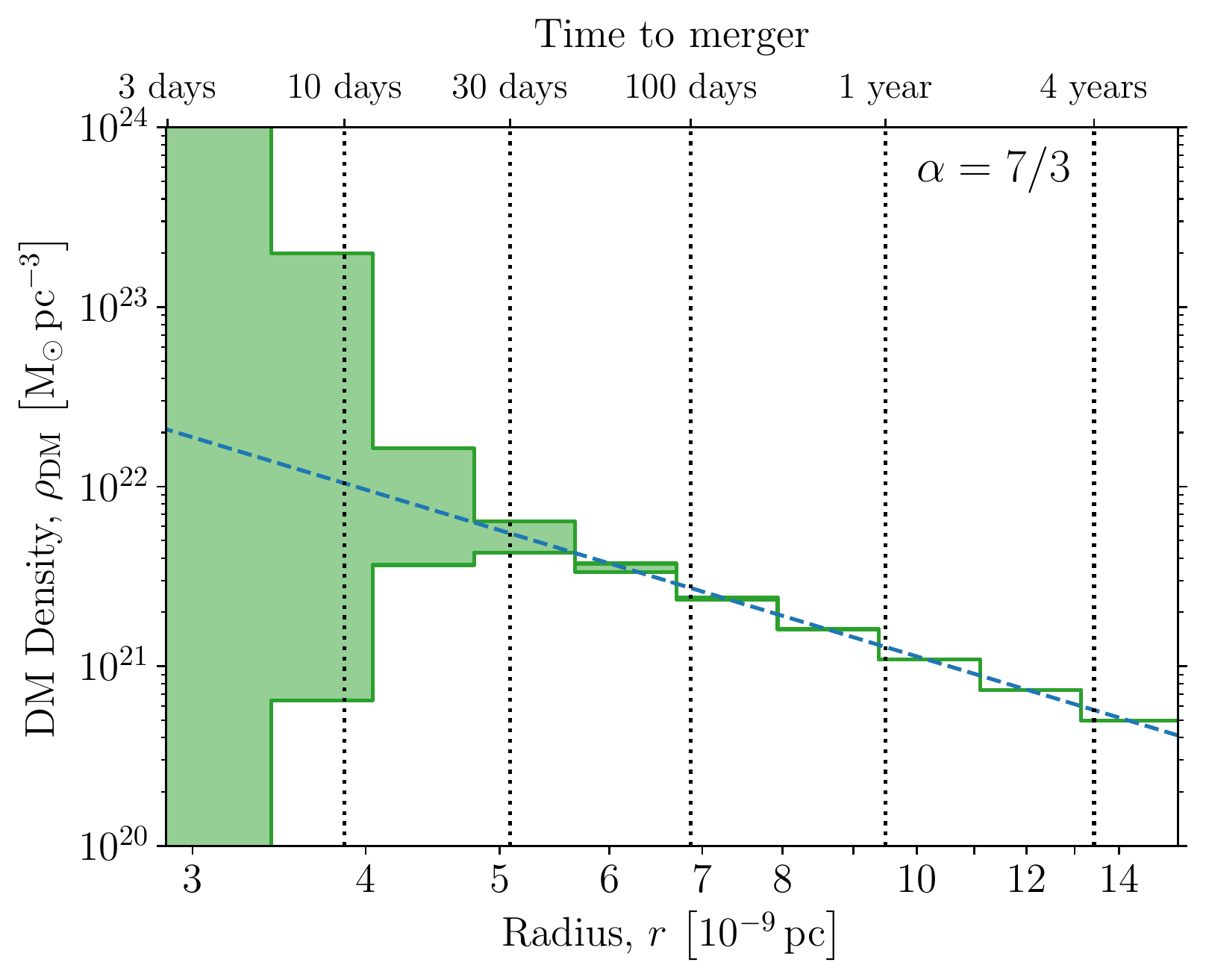}
    \caption{\textbf{Error on the DM density from GW measurements of} $\bm{\alpha}$. Green bands show $1\sigma$ uncertainties on the reconstructed DM density from analysing the GW waveform (for a system at $d=0.01\mathrm{\,Gpc}$, representing a signal-to-noise for LISA of $\sim 92$) over 10 bins in radius (measured from the position of the $10^4 \mathrm{\,M_{\odot}}$ IMBH). The fiducial density profile is shown as a blue dashed line. Along the top axis we also label the approximate time-to-merger in the vacuum case.}
    \label{fig:density}
\end{figure}

Figure~\ref{fig:density} shows the $1\sigma$ uncertainty on the density reconstruction. 
At radii larger than $r \gtrsim 6 \times 10^{-9}\,\mathrm{pc}$, the DM density can be constrained to better than 10\%, but as the separation of the binary decreases the uncertainty on the DM density increases. This is due to three effects; firstly, as the system approaches merger, GW emission (and not DF) begins to dominate the phase evolution of the waveform. Secondly, time spent at a given radius is not evenly distributed, as can be seen in the upper y-axis of Fig.~\ref{fig:density}. Finally, the LISA sensitivity decreases at higher frequencies, weakening the constraining power at small $r$. The phase evolution of the waveform is therefore very sensitive to dynamical friction (and thus the DM density) predominantly when $r$ is large.

While the formation, properties, and survival of DM spikes is not guaranteed \cite{Ullio:2001fb,Merritt:2002vj,Zhao:2005zr,Hannuksela:2019vip}, GW observations can be used to confirm (or disfavor) the presence of a spike in a given system. DM profiles with $\alpha=1.5$ will still induce a detectable phase shift~\cite{Eda:2014kra}, though these shallower slopes will significantly degrade constraints on $\alpha$ (see Supplementary Material~\ref{SM:GW}). In any case, the resulting density constraints can be fed directly into the EM signal calculation, predicting the expected radio emission.

\medskip
\textbf{\textit{Radio Signal ---}} The radio signal arises from resonant axion-photon conversion occurring when the axion mass $m_a$ matches the frequency $\omega_p$ of the plasma surrounding the NS:
\begin{equation}
    \omega_p \approx \sqrt{\frac{4\pi\,\alpha_{\rm EM}\,n_c}{m_c}}\,,
    \label{eq:plasma_frequency}
\end{equation}
with $n_c$ the number density of charged particles with mass $m_c$.
As a concrete example, we consider the Goldreich-Julian model for the NS plasma~\cite{1969ApJ...157..869G},
which in the non-relativistic limit provides
\begin{equation}
    n_c = \frac{2\,\mathbf{\Omega} \cdot \mathbf{B}}{e}\,,
    \label{eq:charge_density}
\end{equation}
where $\Omega= 2\pi/P$ is the angular velocity and $\mathbf{B}$ is the magnetic field, which we consider to be in a dipole configuration with the axis aligned to the rotation axis. Resonant axion-photon conversion then occurs at a specific radial distance from the NS center, which is given by~\cite{Hook:2018iia}
\begin{eqnarray}
    r_c &\simeq& 58\,{\rm km}\left|3\cos^2\theta-1\right|^{1/3} \left(\frac{r_{\rm NS}}{10\,{\rm km}}\right) \times \nonumber\\
    &&\left[\frac{B_0}{10^{12}\,{\rm G}}\frac{10\,{\rm s}}{P}\left(\frac{10^{-6}\,{\rm eV}}{m_a}\right)^2\right]^{1/3}\,,
    \label{eq:r_c}
\end{eqnarray}
where $\theta$ is the polar angle with respect to the rotation axis. Equation~\eqref{eq:r_c} is obtained by setting $\omega_p=m_a / 2\pi$ and considering electrons/positions plasma ($m_c = m_e$). 

Following Ref.~\cite{Hook:2018iia}, the radiated power is given by
\begin{equation}
    \frac{\mathrm{d}\mathcal{P}}{\mathrm{d}\Omega} \sim 2\times p_{a\gamma}\,\rho_{\text{DM}}(r_c)\,v_c\,r_c^2\,,
    \label{eq:power}
\end{equation}
where $\rho_{\rm DM}(r_c)$ and $v_c$ are the DM density and velocity at the conversion radius
. The energy transfer function $p_{a\gamma}$ is obtained using the WKB and stationary phase approximations to give
\begin{equation}
    p_{a\gamma} \sim \frac{\pi}{12}\frac{g_{a\gamma\gamma}^2\,B_0^2\,\,r_{\rm NS}}{m_a}\left(\frac{r_{\rm NS}}{r_c}\right)^5\left(3\cos^2\theta+1\right)\,,
    \label{eq:prob_conv}
\end{equation}
where $g_{a\gamma\gamma}$ is the strength of the coupling that leads to axion-photon conversion through the interaction $\mathcal{L} = - g_{a\gamma\gamma} \,a \,\mathbf{E}\cdot\mathbf{B}/4$.

We use Eddington's formula to calculate the 
phase-space distribution of the DM in the BH frame \cite{2008gady.book.....B,Catena:2011kv}, assuming isotropy and spherical symmetry (see Supplementary Material~\ref{SM:velocity_dist} for more details). This distribution $f(\mathcal{E})$ depends on the relative energy $\mathcal{E} = \Psi(r) - \frac{1}{2}v^2$ and the gravitational potential $\Psi(r) = \Phi_0 - \Phi(r)$, relative to the potential at the mini-spike boundary, $\Phi_0$. For $r \lesssim 10^{-8}$ pc (the point at which the GW signal would become observable) the enclosed mass is dominated by the BH and we therefore neglect the contribution of the mini-spike to the relative potential: $\Psi = \Psi_\mathrm{BH} = G_N \, M_\mathrm{BH}/r$. In this case, we find $f(\mathcal{E}) \propto \mathcal{E}^{\alpha-3/2}$ (for $\mathcal{E} > 0$).
 
 Nearby DM particles are accelerated under gravity as they infall toward the NS. Particles with initial velocity $v$ reach velocity $\sqrt{v^2 + 2\Psi_\mathrm{NS}}$ at the conversion radius, where the NS potential is $\Psi_\mathrm{NS} = G_N M_\mathrm{NS}/r_c$. Applying Liouville's theorem~\cite{Liouville:1838zza}, we find the DM density at the conversion radius as,
\begin{equation}
\begin{split}
    \rho_{\text{DM}}(r_c) &= \sqrt{\frac{2}{\pi}}\frac{\rho_{\mathrm{sp}}r_{\mathrm{sp}}{}^{\alpha}}{(G_N M_{\mathrm{BH}})^{\alpha}}\frac{ \alpha(\alpha-1)\Gamma(\alpha-1)}{\Gamma(\alpha-\frac{1}{2})}\times \\
    &\quad\int^{v_{\mathrm{max}}}_{v_{\mathrm{min}}} \left[\Psi_\mathrm{BH} + \Psi_\mathrm{NS} - \frac{v^2}{2} \right]^{\alpha-\frac{3}{2}}v^2\,  \mathrm{d}v,
\end{split}
\end{equation}
where $v_\mathrm{min} = \sqrt{2\Psi_\mathrm{NS}}$ and $v_\mathrm{max} = \sqrt { 2 \left( \Psi _ { \mathrm { BH } } + \Psi _ { \mathrm { NS } }\right)}$.
We assume that the amplitude of the radiated power is dominated by the peak of the velocity distribution:\footnote{We do not consider the boost to the NS frame since the NS orbital velocity is subdominant with respect to the DM peak velocity.} 
\begin{equation}
     v_c^2 \sim \frac{2G_N M_{\mathrm{BH}}}{r}\left[\alpha - \frac{1}{2}\right]^{-1} + \frac{2G_N M_{\mathrm{NS}}}{r_c}\,.
     \label{eq:v_c}
\end{equation}
Finally, the flux density of the radio signal is given by
\begin{equation}
    S = \frac{1}{\mathcal{B}\,d^2} \frac{\mathrm{d}\mathcal{P}}{\mathrm{d}\Omega} \,,
    \label{eq:flux_density}
\end{equation}
where $d$ is the distance to the system and $\mathcal{B}$ is the signal bandwidth (calculated as the 90\% containment region of the DM velocity distribution far from the NS). Given the central frequency $f$ of the radio signal, we find $\mathcal{B}/f$ to be 0.06 and 0.12 at an orbit of $r=6\times 10^{-9}~\mathrm{pc}$ and $r=3\times 10^{-9}~\mathrm{pc}$, respectively.
\begin{figure}[t!]
    \centering
    \includegraphics[width=1.0\linewidth]{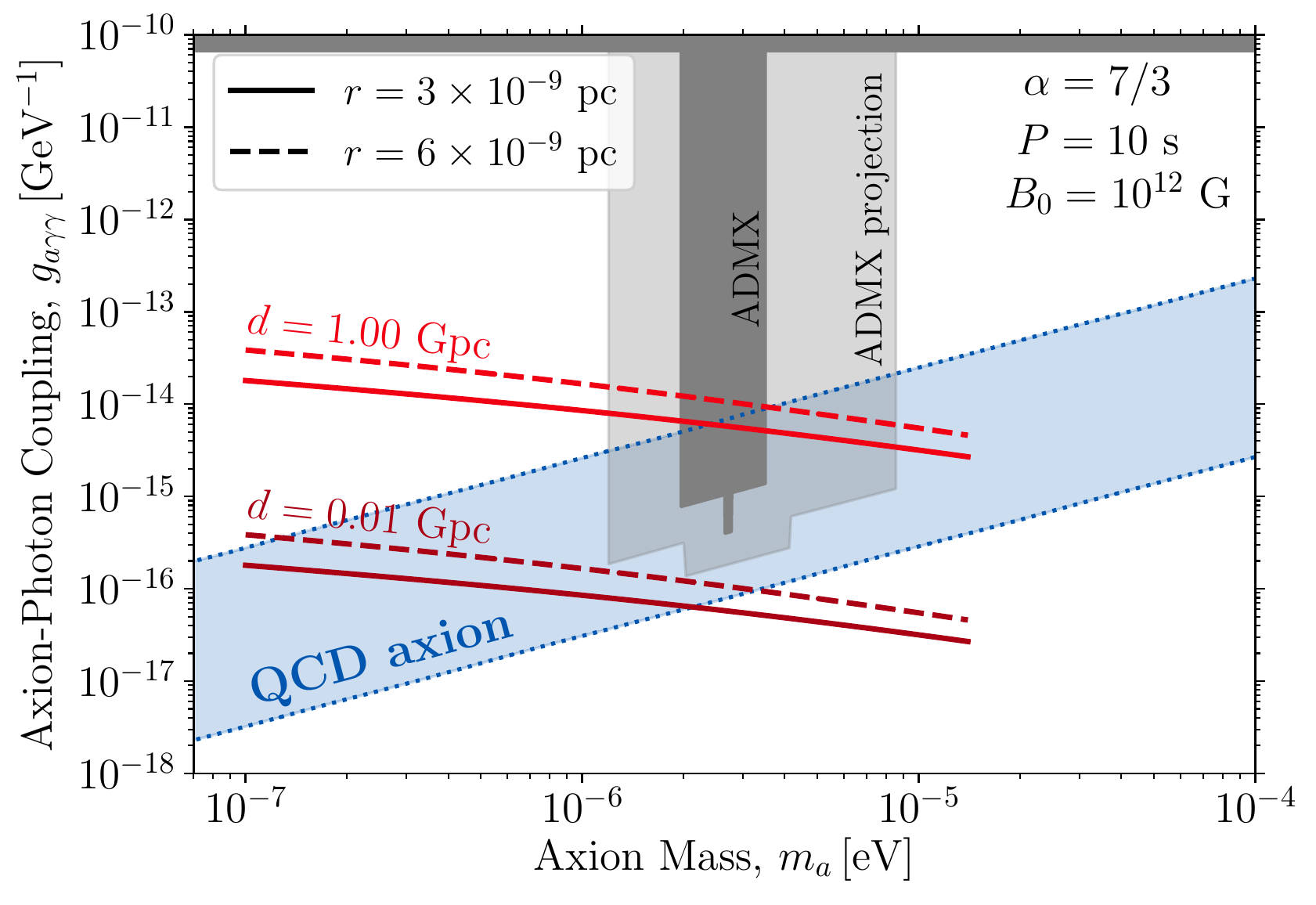}
    \caption{\textbf{Projected reach in axion-photon coupling from radio observations.} SKA2 sensitivity 
    (100 hours) to the axion-photon coupling for different orbital separations $r$ and IMBH-DM-NS distances $d$, assuming $\theta=90^\circ$.
    The QCD axion parameter space is represented by the blue band, while the vertical and horizontal gray bands show the ADMX~\cite{Asztalos:2009yp,Du:2018uak} and CAST~\cite{Anastassopoulos:2017ftl} limits, respectively.}
    \label{fig:limit}
\end{figure}

Figure~\ref{fig:limit} shows the projected reach of the future SKA2 telescope in the axion parameter space, obtained by considering the minimum detectable flux density $S$ which provides a signal-to-noise ratio (SNR) equal to one.\footnote{An SNR of one was chosen to directly compare with Ref.~\cite{Hook:2018iia}. Increasing the required SNR for detection leads to a corresponding increase in the couplings we can probe as $g_{a\gamma\gamma} \propto \sqrt{\mathrm{SNR}}$.} In particular, for a radio telescope,
\begin{equation}
    {\rm SNR} =  \frac{S}{\rm SEFD} \sqrt{n_{\rm pol}\,\mathcal{B}\,\Delta t_{\rm obs}} \,,
    \label{eq:snr}
\end{equation}
where $n_{\rm pol} = 2$ is the number of polarizations,
$\rm SEFD = 0.098$ is the SKA2 system-equivalent flux density as estimated in Ref.~\cite{Safdi:2018oeu}
and we assume an observation time of $\Delta t_{\rm obs} = 100$ hours. This is roughly the time spent by the system from the closest orbit we consider ($r = 3 \times 10^{-9}$~pc) until merger.
We have assumed that the sensitivity is limited by the detector's thermal noise.
Since IMBHs have not yet been conclusively detected, it is thought that they do not have an appreciable accretion disk, meaning that there is unlikely to be any background radio emission during the inspiral.
The sensitivity curves are valid in the mass range $10^{-7} \mathrm{\, eV}\leq m_a \leq  1.4 \times 10^{-5} \mathrm{\, eV}$. The lower limit is set by the lower cut-off frequency potentially probed by SKA,\footnote{We note that for $m_a > 10^{-7} \mathrm{\, eV}$ the corresponding axion Compton wavelength is smaller than $0.01 \mathrm{\, km}$, three orders of magnitude smaller than the size of the NS. This allows us to neglect 
quantum effects when computing both signatures.} while the upper limit comes from the requirement that axion-photon conversion occurs outside the NS, $r_c \geq r_{\rm NS}$.

Figure~\ref{fig:limit} shows that a crucial parameter is the distance of the IMBH-DM-NS system, since the flux density depends on its inverse square. On the other hand, the sensitivity does not strongly depend on the BH-NS separation $r$. Radio observations taken when $r \sim 3 \times 10^{-9}\,\mathrm{pc}$ (solid lines) yield sensitivities to $g_{a\gamma\gamma}$ which are roughly a factor of 2 stronger than for $r \sim 6 \times 10^{-9}\,\mathrm{pc}$ (dashed lines). In Fig.~\ref{fig:limit}, we have fixed $\rho_\mathrm{DM}$ to the fiducial density profile. However, as we saw in Fig.~\ref{fig:density}, the DM density is likely to be more poorly constrained at smaller radii, making the radio sensitivity at large $r$ substantially more robust (though not substantially weaker).

\medskip
\textbf{\textit{Discussion ---}} With a sufficiently nearby detection of an IMBH-DM-NS system, it will be possible to probe the parameter space of QCD axion DM. We find roughly a $0.05\%$ probability of a detection closer than $d = 0.01 \,\mathrm{Gpc}$ over 10 years, using predicted LISA detection rates for such systems \cite{Fragione:2017blf}   (though these typically come with large uncertainties). Instead, out to $d = 1 \,\mathrm{Gpc}$, we expect a few tens of detections per year.

We emphasise that setting upper limits on $g_{a\gamma\gamma}$ from the \textit{non-detection} of a radio signal is hampered by uncertainties in the individual NS properties and magnetosphere modelling~(see Supplementary Material~\ref{SM:NSparam}). Detecting and studying a larger population of such systems would perhaps allow for robust limit-setting, through modelling of the expected properties of the NS population. Nevertheless, a joint GW+EM detection is within reach of upcoming experiments and would be a striking confirmation of axion Dark Matter. GW observations can provide constraints on the DM density around BHs, as in Fig.~\ref{fig:density}, with the better estimation of the density at larger separations reducing uncertainties on the expected radio signal. 

Above around $m_a \sim 10^{-6}\,\mathrm{eV}$, these broadband sensitivities would be complementary to current and proposed axion haloscope experiments \cite{Asztalos:2009yp,Du:2018uak,Brubaker:2016ktl,TheMADMAXWorkingGroup:2016hpc,Zhong:2018rsr,Brun:2019lyf} (some of which are plotted in Fig.~\ref{fig:limit}). These are sensitive to the density of DM local to Earth, which carries its own uncertainties \cite{Read:2014qva}. Such uncertainties could be mitigated in our scenario by combining information from GW and radio emission. Multi-messenger observations of Black Hole - Dark Matter - Neutron Star systems therefore have the potential to detect QCD axion Dark Matter for masses between $10^{-7}\,\mathrm{eV}$ and $10^{-5}\,\mathrm{eV}$.

\acknowledgments{
We thank David Nichols, Tanja Hinderer, Mikael Leroy, and Gianfranco Bertone for fruitful discussions. Finally, we thank the python scientific computing packages numpy \cite{numpy} and scipy \cite{scipy}. This research is funded by NWO through the VIDI research program ``Probing the Genesis of Dark Matter" (680-47-532; TE, CW).}

\bibliography{main}

\clearpage
\newpage
\maketitle
\onecolumngrid
\begin{center}
\textbf{\large A Unique Multi-Messenger Signal of QCD Axion Dark Matter} \\ 
\vspace{0.1in}
{ \it \large Supplementary Material}\\ 
\vspace{0.05in}
{Thomas D. P. Edwards, \ Marco Chianese, \ Bradley J. Kavanagh, Samaya M. Nissanke, and \ Christoph Weniger}
\end{center}
\onecolumngrid
\setcounter{equation}{0}
\setcounter{figure}{0}
\setcounter{table}{0}
\setcounter{section}{0}
\setcounter{page}{1}
\makeatletter
\renewcommand{\theequation}{S\arabic{equation}}
\renewcommand{\thefigure}{S\arabic{figure}}

This Supplementary Material is organized as follows: In App.~\ref{SM:GW} we discuss both the gravitational wave and radio signals dependence on the Dark Matter (DM) spike parameters. Appendix~\ref{SM:velocity_dist} discusses the velocity distribution of the DM, highlighting its limitations and how it can be addressed in future work. Finally, App.~\ref{SM:NSparam} discusses the dependence of the radio signal on the Neutron Star (NS) parameters. Here, we also speculate about the amplification of the radio signal if the neutron star had magnetic field strengths up to $10^{15} \,\mathrm{G}$ or spin periods down to $0.1~{\rm s}$.

\section{Gravitational Waves and Spike Dependence}\label{SM:GW}

\begin{figure}[tbh!]
    \centering
    \includegraphics[width=0.5\linewidth]{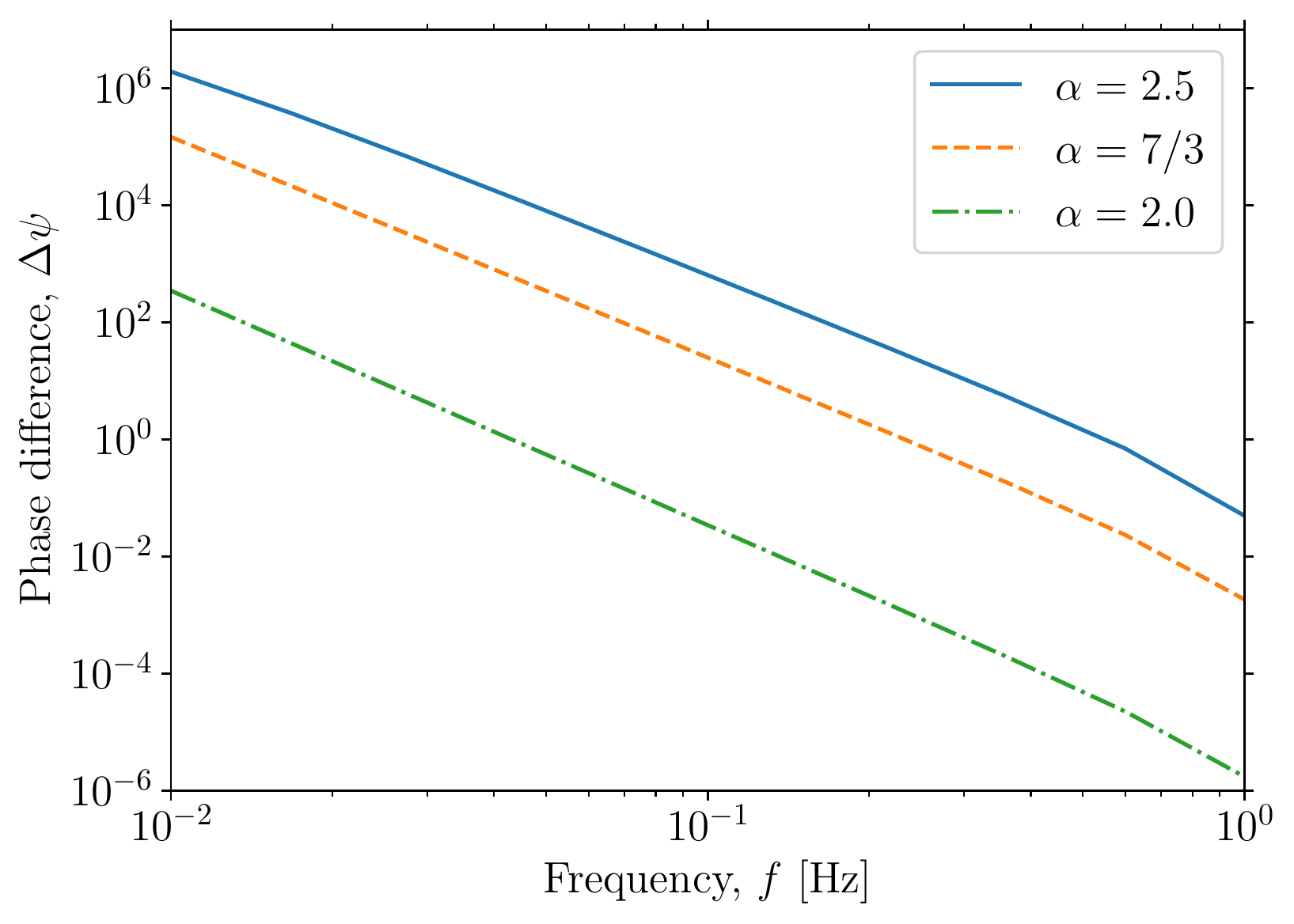}
    \caption{\textbf{Phase difference between vacuum and DM spike inspiral.} We show the difference in the phase evolution of the IMBH-DM-NS system compared to a vacuum IMBH-NS inspiral for $\alpha = \{2.0, \, 7/3,\, 2.5\}$. As $\alpha$ is increased, the phase difference becomes larger. Similarly, the phase difference continues to be significant for higher frequencies when $\alpha>7/3$. This persistent phase shift for large $\alpha$ is reflected as tighter constraints on the DM density, as seen in Fig.~\ref{fig:rho_a}.}
    \label{fig:Phasedif}
\end{figure}

The phase difference between a vacuum inspiral and the one considered here is given by,
\begin{equation}
    \Delta\psi = \tilde{\phi} - \phi\,,
\end{equation}
where $\phi = -\frac{3}{4}\left(8\pi G_N \mathcal { M }_c f/c^3\right)^{-5/3}$ is the Newtonian vacuum phase evolution and $\tilde{\phi}$ is the phase evolution including dynamical friction, as given by Eq.~(28) of Ref.~\cite{Eda:2014kra}. The phase evolution of the gravitational wave signal provides a fundamental insight into the dynamics of the binary system. As shown in Fig.~\ref{fig:Phasedif}, the presence of a DM spike (with $\alpha>2.0$) produces a considerable phase shift when compared to the evolution of a vacuum inspiral. Again for $\alpha>2.0$, the specific phase evolution of any particular system can therefore be used to constrain $\alpha$ to high precision. For our baseline scenario of $\alpha=7/3$, constraints on $\alpha$ correspond to an error on the DM density of $\mathcal{O}(0.01\%)$ at $r\approx 1.3\times 10^{-8}\,\mathrm{pc}$. For $\alpha<2.0$, the phase difference becomes increasingly difficult to probe. Assuming the masses of the two objects can be independently measured to high precision, the constraint on $\alpha$ provides a direct constraint on the DM density local to the position of the NS \cite{Kavanagh:2020cfn}. We do not account for errors on the overall normalisation of the DM density profile directly. The normalisation can also be measured, though it is degenerate with $M _ \mathrm{ NS }$ and $M _ \mathrm{ BH }$. To resolve the individual masses, higher order effects close to merger need to be accounted for. The errors associated with the individual mass determinations may dominate the error on the DM density normalisation at larger radii, but this is beyond the scope of the paper. We will address this in future work.

\begin{figure}[t!]
    \centering
    \includegraphics[width=0.49\linewidth]{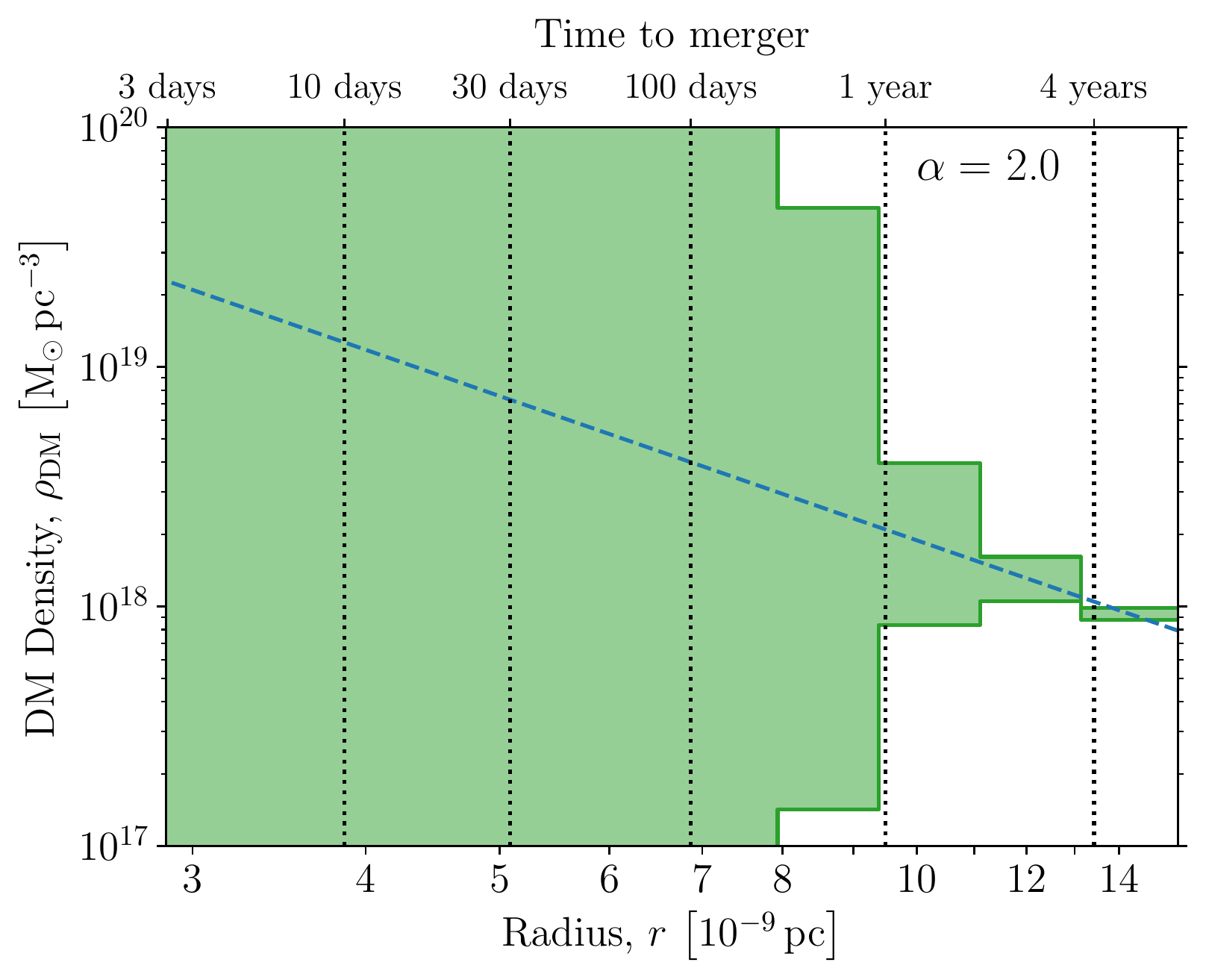}
    \includegraphics[width=0.49\linewidth]{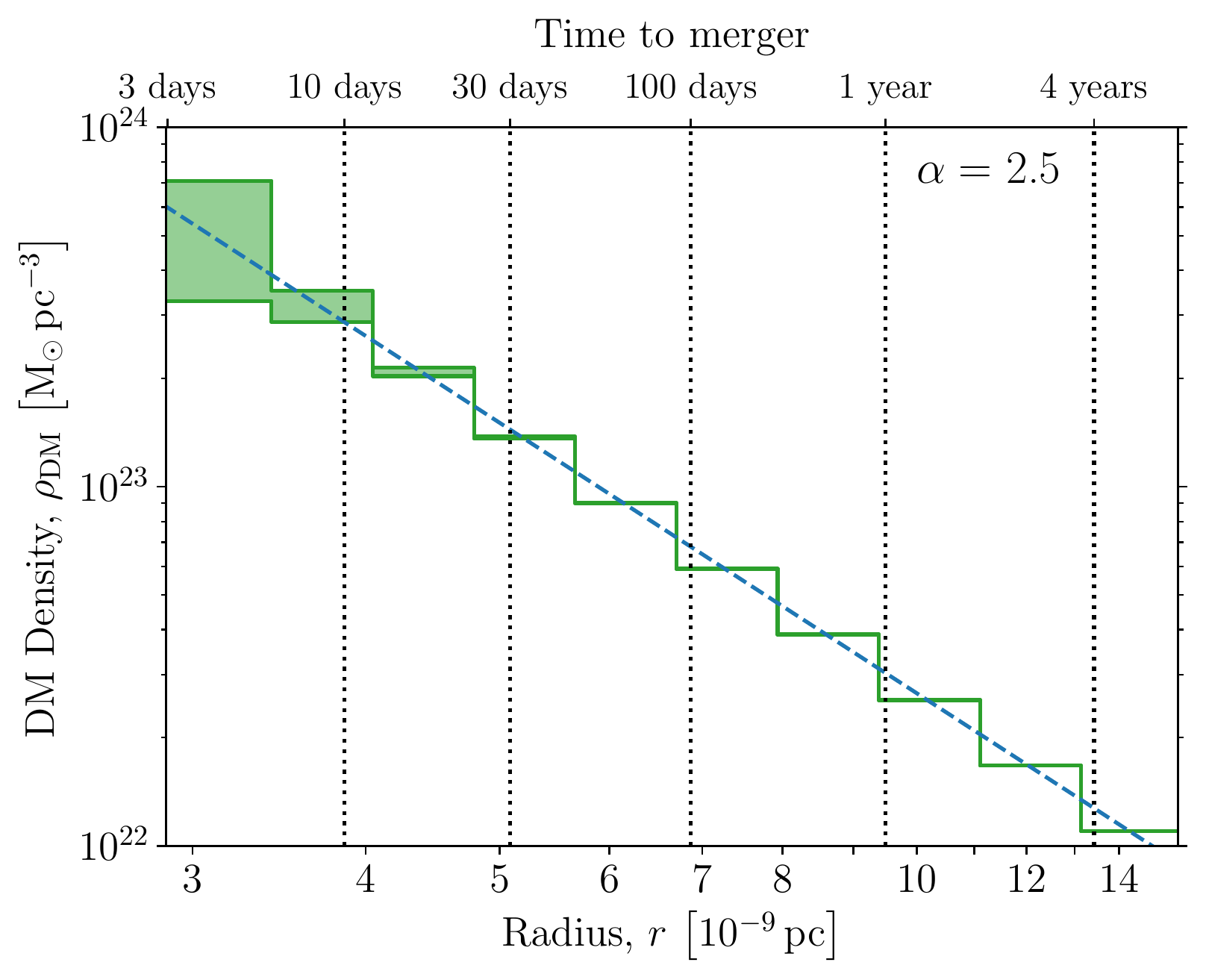}
    \caption{\textbf{Error on the DM density from GW measurements of} $\bm{\alpha}$. Green bands show the $1\sigma$ uncertainties on the reconstructed DM density from analysing the GW waveform (for a system at $d=0.01\mathrm{\,Gpc}$) over 10 bins in radius (measured from the position of the $10^4 \mathrm{\,M_{\odot}}$ IMBH). The fiducial density profiles are shown as a blue dashed line with $\alpha = \{2.0, \, 7/3,\, 2.5\}$ on the left and right respectively. Along the top axis we also label the approximate time-to-merger as a function of radius in the vacuum case.}
    \label{fig:rho_a}
\end{figure}

Figure~\ref{fig:rho_a} shows the constraint on the DM density from the GW signal, as described in the main text. The left and right panels show the constraint for $\alpha=2.0$ and $\alpha=2.5$, respectively. The error bars become larger for lower values of $\alpha$; this can easily be understood from Fig.~\ref{fig:Phasedif}. As the inspiral progresses, the GW frequency becomes larger (equivalently, the radius decreases). Similarly, the phase difference becomes ever smaller, gradually approaching the vacuum inspiral phase evolution and therefore providing no probe of the DM density. As $\alpha$ is increased, the large phase differences persist further into the inspiral, allowing one to probe the DM density closer to the IMBH.

\begin{figure}[t!]
    \centering
    \includegraphics[width=0.5\linewidth]{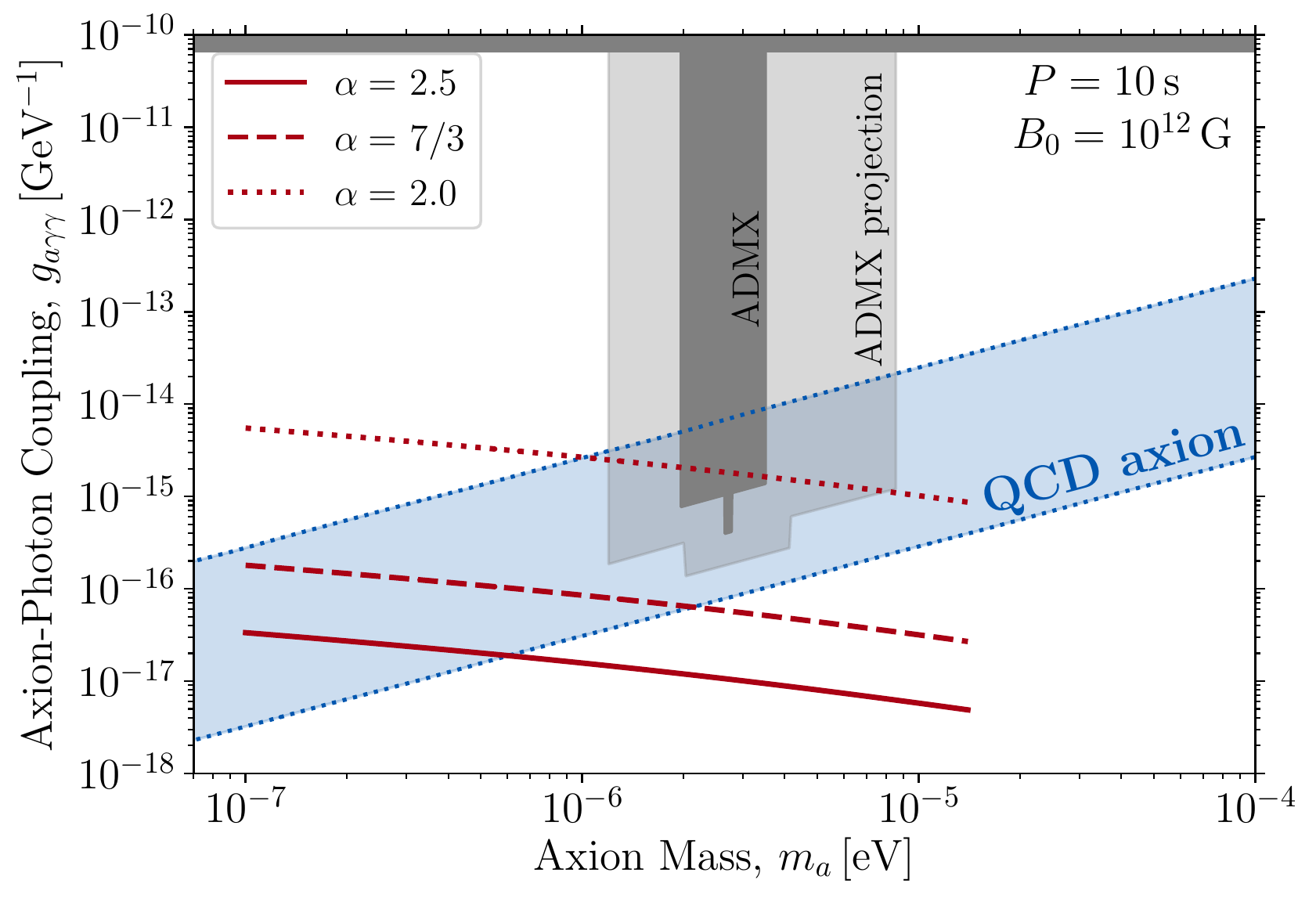}
    \caption{\textbf{Projected reach in axion-photon coupling from radio observations.} Sensitivity curves of the SKA2 telescope (100 hours of observation) to the axion-photon coupling as a function of the axion mass for $\alpha = \{2.0, \, 7/3,\, 2.5\}$. Here, we assume a radial separation of $r=3\times 10^{-9}$\,pc, $d=0.01$\,Gpc for the distance to the system, $B_0 = 10^{12}$\,G for the NS magnetic field, $P=10$\,s as the NS spin period, and $\theta=90^\circ$. The predicted range of parameters for the QCD axion are represented by the blue band, while the vertical and horizontal gray bands show the ADMX~\cite{Asztalos:2009yp,Du:2018uak} and CAST~\cite{Anastassopoulos:2017ftl} limits, respectively.}
    \label{fig:limit_a}
\end{figure}

Finally, in Fig.~\ref{fig:limit_a} we present the radio sensitivity for $\alpha = \{2.0, \, 7/3,\, 2.5\}$. As expected, the varying density, as shown in Fig.~\ref{fig:rho_a}, amplifies or decreases the density close the IMBH. For $\alpha=2.0$, it is still possible to probe a small range of the QCD axion parameter space, although the constraint on the DM density becomes significantly worse (see left panel of Fig.~\ref{fig:rho_a}). When $\alpha=2.5$, the density can be constrained extremely well down to small radii. The density is also increased by an order of magnitude compared to the $\alpha=7/3$ scenario, subsequently increasing the sensitivity by a similar amount.

\section{Dark Matter Velocity Distribution}\label{SM:velocity_dist}

We assume that the distribution of DM around the central black hole is spherically symmetric and that the velocity distribution of DM particles is isotropic. In this case, we can calculate the DM distribution function using Eddington's Inversion Formula:
\begin{align}
    f ( \mathcal { E } ) = \frac { 1 } { \sqrt { 8 } \pi ^ { 2 } } \int _ { 0 } ^ { \mathcal { E } } \frac { \mathrm { d } \Psi } { \sqrt { \mathcal { E } - \Psi } } \frac { \mathrm { d } ^ { 2 } \rho } { \mathrm { d } \Psi ^ { 2 } } \,.
\label{eq:Eddington}
\end{align}
Here, $\rho(r)$ is the density profile of the DM particles, while $\Psi(r)$ is the total gravitational potential, which in general includes a contribution from both the central mass and the mass enclosed in the DM halo. However, at small radii, the enclosed DM mass is small and we can typically neglect the contribution of the DM halo itself to the gravitational potential. We thus write $\Psi(r) = G_N\,M_\mathrm{BH}/r$ and re-express the density in terms of the potential:
\begin{align}
    \rho(\Psi) &= \rho_\mathrm{sp} \left(\frac{r_\mathrm{sp}}{G_N\,M_\mathrm{BH}}\right)^{\alpha} \Psi^\alpha\,.
\end{align}
We therefore find:
\begin{align}
\begin{split}
    f ( \mathcal { E } ) &= \frac { \alpha(\alpha-1)} { \sqrt { 8 } \pi ^ { 2 } }\rho_\mathrm{sp}\left(\frac{r_\mathrm{sp}}{G_N \,M_\mathrm{BH}}\right)^\alpha  \int _ { 0 } ^ { \mathcal { E } } \Psi^{\alpha-2}\,\frac { \mathrm { d } \Psi } { \sqrt { \mathcal { E } - \Psi } }  \\
    &= \frac { \alpha(\alpha-1)} { (2\pi)^{3/2} }\rho_\mathrm{sp}\left(\frac{r_\mathrm{sp}}{G_N \,M_\mathrm{BH}}\right)^\alpha  \frac{\Gamma(\alpha -1)}{\Gamma(\alpha -\frac{1}{2})} \mathcal{E}^{\alpha - 3/2}\,.
\label{eq:f_E}
\end{split}
\end{align}

\begin{figure}[t!]
    \centering
    \includegraphics[width=0.50\linewidth]{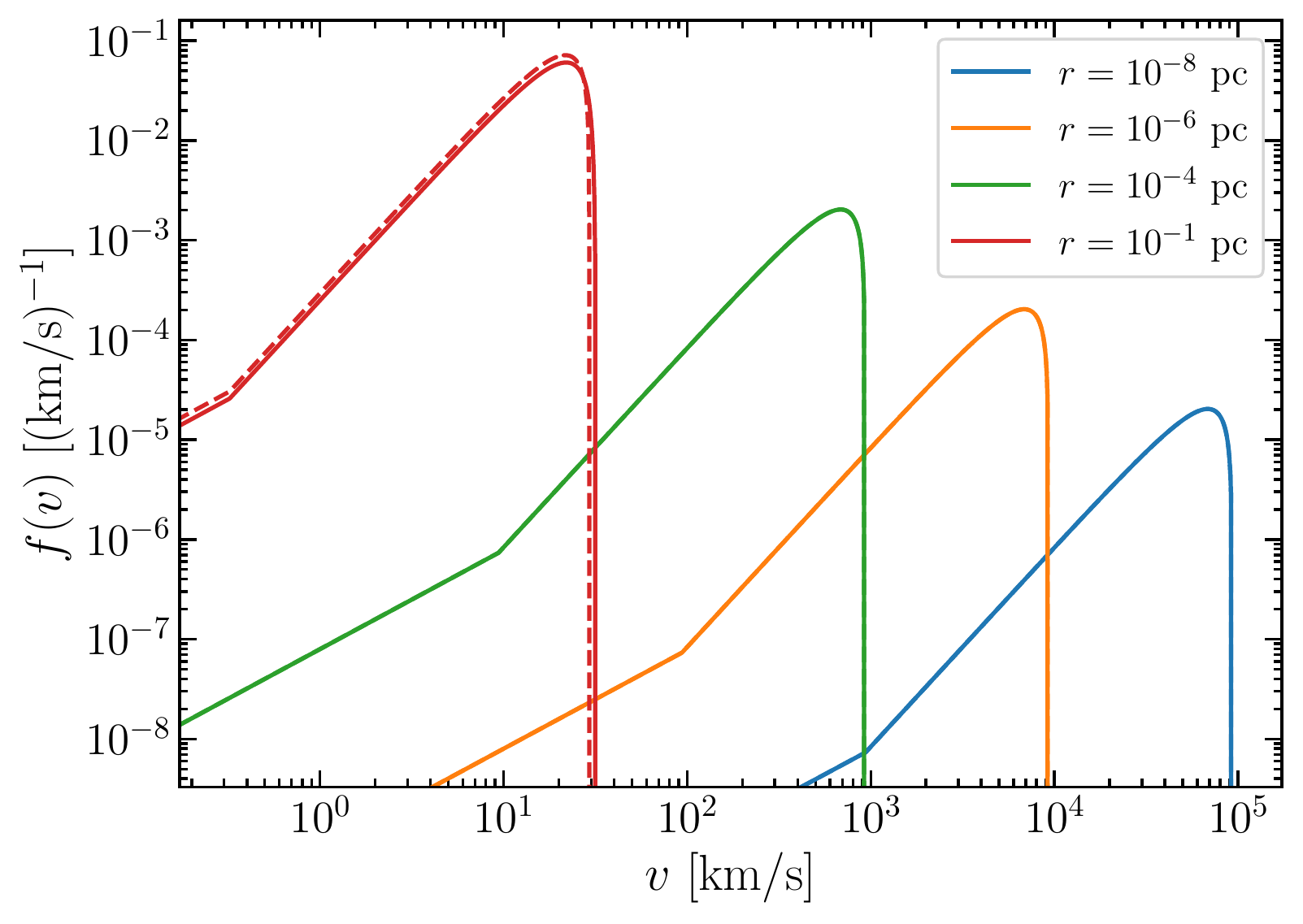}\caption{\textbf{Dark Matter Speed distributions.} DM speed distribution derived from the Eddington Inversion formula, Eq.~\eqref{eq:Eddington}, at different distances $r$ from the central IMBH, $M_\mathrm{IMBH} = 10^{4}\,M_\odot$. Solid lines show the full calculation accounting for the potential due to the DM halo itself, while dashed lines show the approximate result, Eq.~\eqref{eq:f_approx}, including only the potential due to the IMBH. Here, we assume $\alpha = 7/3$.}
    \label{fig:DMdistribution}
\end{figure}

The DM speed distribution at a radius $r$ is then given by
\begin{align}
f(v|r) &= 4\pi v^2 \frac{f(\Psi(r) - \frac{1}{2}v^2)}{\rho(r)} =  \frac{4}{\sqrt{\pi}}\frac{\Gamma\left( \alpha + 1 \right)}{\Gamma\left(\alpha - \frac{1}{2}\right)}\frac{v^2}{v_\mathrm{max}^{2\alpha}} \left(v_\mathrm{max}^2 - v^2\right)^{\alpha - 3/2}\,.
\label{eq:f_approx}
\end{align}
Here, we have defined $v_\mathrm{max} = v_\mathrm{max}(r) = \sqrt{2\Psi(r)}$, and we set the speed distribution to zero for $v > v_\mathrm{max}(r)$.  With this definition, the speed distribution is normalised to one at any given radius:
\begin{equation}
 \int_0^{v_\mathrm{max}(r)} f(v|r) \,\mathrm{d}v = 1\,.
\end{equation}
The peak velocity (far from the NS) is obtained by setting $\partial f(v|r)/\partial v = 0 $, giving:
\begin{equation}
    v_\mathrm{peak}^2 = v_\mathrm{max}^2 \left[\alpha-\frac{1}{2}\right]^{-1} = \frac{2 G_N M_\mathrm{BH}}{r}\left[\alpha-\frac{1}{2}\right]^{-1}\,.
\end{equation}

In Fig.~\ref{fig:DMdistribution}, we show the DM speed distribution at several radii $r$, assuming $\alpha = 7/3$ and $M_\mathrm{IMBH} = 10^4\,M_\odot$. Solid lines show the speed distribution derived from a full numerical calculation of $f(\mathcal{E})$ (using Eq.~\eqref{eq:f_E} and including self-consistently the potential due to the DM halo). Dashed lines show the approximate speed distribution given in Eq.~\eqref{eq:f_approx} (neglecting the potential of the DM halo itself). We see that in all cases of interest to us, $r \lesssim 10^{-8} \,\mathrm{pc}$, Eq.~\eqref{eq:f_approx} provides an excellent approximation to the full expression.

We note that as $r \rightarrow r_\mathrm{ISCO}$, the maximum DM speed tends towards the speed of light. As we discuss in the main text, the dominant effect from dynamical friction typically occurs at larger radii, where the DM speeds are lower and the non-relativistic formalism should apply. However, at a radius $r = 3 \times 10^{-9}\,\mathrm{pc}$, the maximum DM speed is $v_\mathrm{max} \approx 1.7 \times 10^{5}\,\mathrm{km/s} \approx 0.56 \,c$. Adding also the infall velocity toward the conversion radius, the DM particles are accelerated up to $\sim 0.8 \,c$. This suggests that towards the end of the inspiral, our non-relativistic formalism would over-estimate the speeds which can be reached by the DM particles. However, even at $r = 3 \times 10^{-9}\,\mathrm{pc}$, the DM speeds are still only mildly relativistic ($\gamma \sim 1.7$), suggesting that this should be a small effect. We leave a more detailed analysis -- including relativistic effects, boosting in the NS rest-frame and anisotropy of the infalling DM flux -- to future work.

Finally, we have checked that the DM halo should survive the merger itself; the work done by dynamical friction during the five year inspiral is only a few percent of the total gravitational binding energy of the halo. A more detailed study of feedback on the DM halo in different systems is in preparation \cite{Kavanagh:2020cfn}.

\section{Neutron Star Parameters}\label{SM:NSparam}

In the Goldreich-Julian model for the NS magnetosphere~\cite{1969ApJ...157..869G}, the amplitude of the radio signal (produced by the resonant axion-photon conversion) is completely determined by the magnetic field strength at the NS poles $B_0$ and the spin period $P$. In particular, they determine the number density of charged particles around the NS (Eq.~\eqref{eq:charge_density}), and consequently the plasma frequency~\eqref{eq:plasma_frequency} and the conversion radius~\eqref{eq:r_c}. By plugging Eq.~\eqref{eq:prob_conv} into Eq.~\eqref{eq:power}, we find the scaling relatation of the radiated power with respect to $B_0$ and $P$ to be
\begin{equation}
    \frac{\mathrm{d}\mathcal{P}}{\mathrm{d}\Omega} \propto  B_0 \, P\left(\frac{3\cos^2\theta + 1}{\left|3\cos^2\theta - 1\right|}\right)\left[g_{a\gamma\gamma}^2\,m_a\, \rho_{\rm DM}(r_c) \, v_c\right]\,.
    \label{eq:power_dep}
\end{equation}
where the quantities in the squared parentheses are almost independent of the NS parameters. Hence, for a given axion mass, the larger the NS magnetic field and spin period, the larger the radiated power. Furthermore, the radiated power can be significantly larger for $\cos\theta = 1/\sqrt3$. We note that by neglecting the second term in the expression~\eqref{eq:v_c} for the velocity $v_c$, one can show that the minimum detectable axion-photon coupling from Eq.~\eqref{eq:snr} scales with the axion mass as $g_{a\gamma\gamma}^{\rm min} \sim m_a^{-1/2}$.

\begin{figure}[t!]
    \centering
    \includegraphics[width=0.49\linewidth]{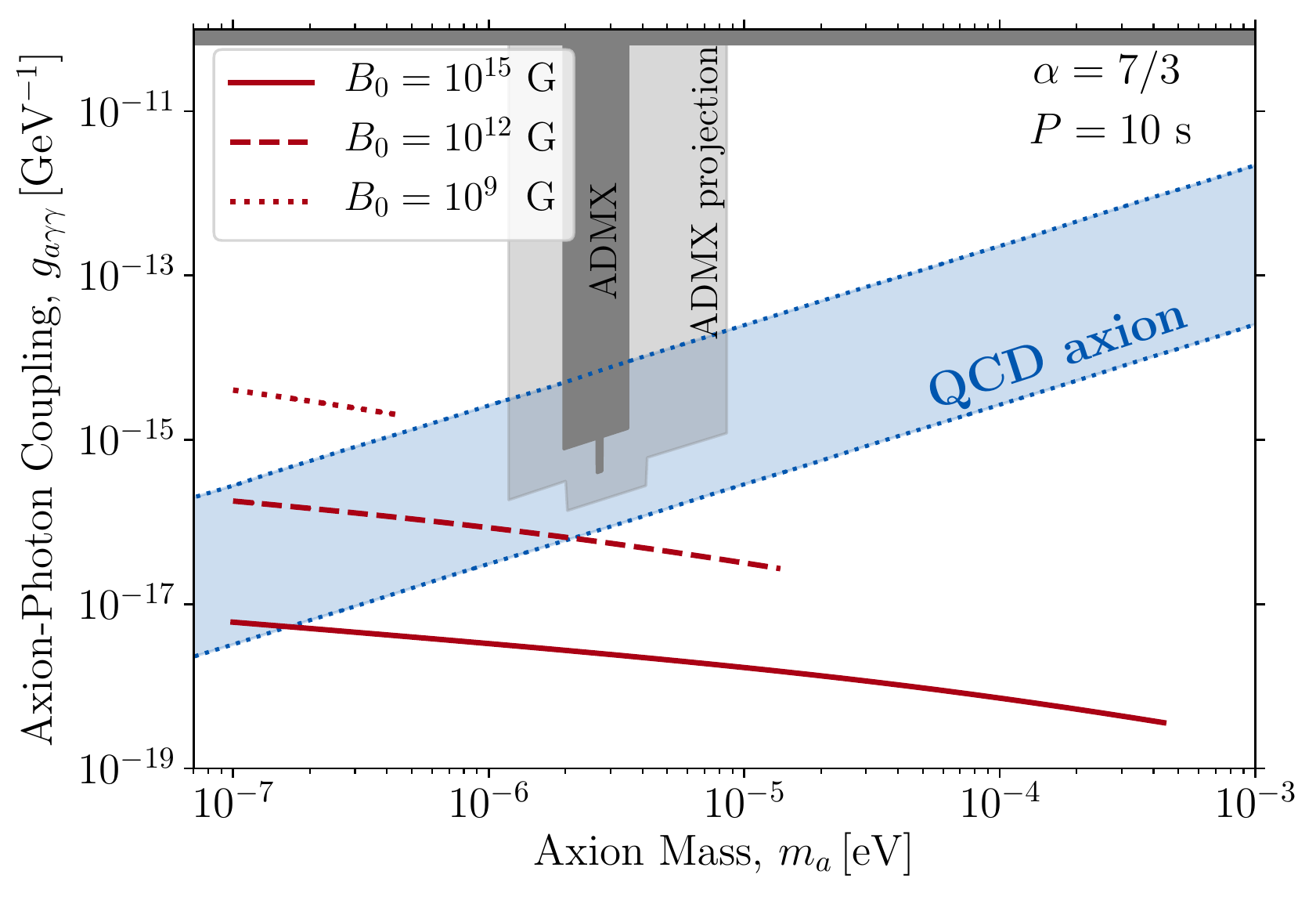}
    \includegraphics[width=0.49\linewidth]{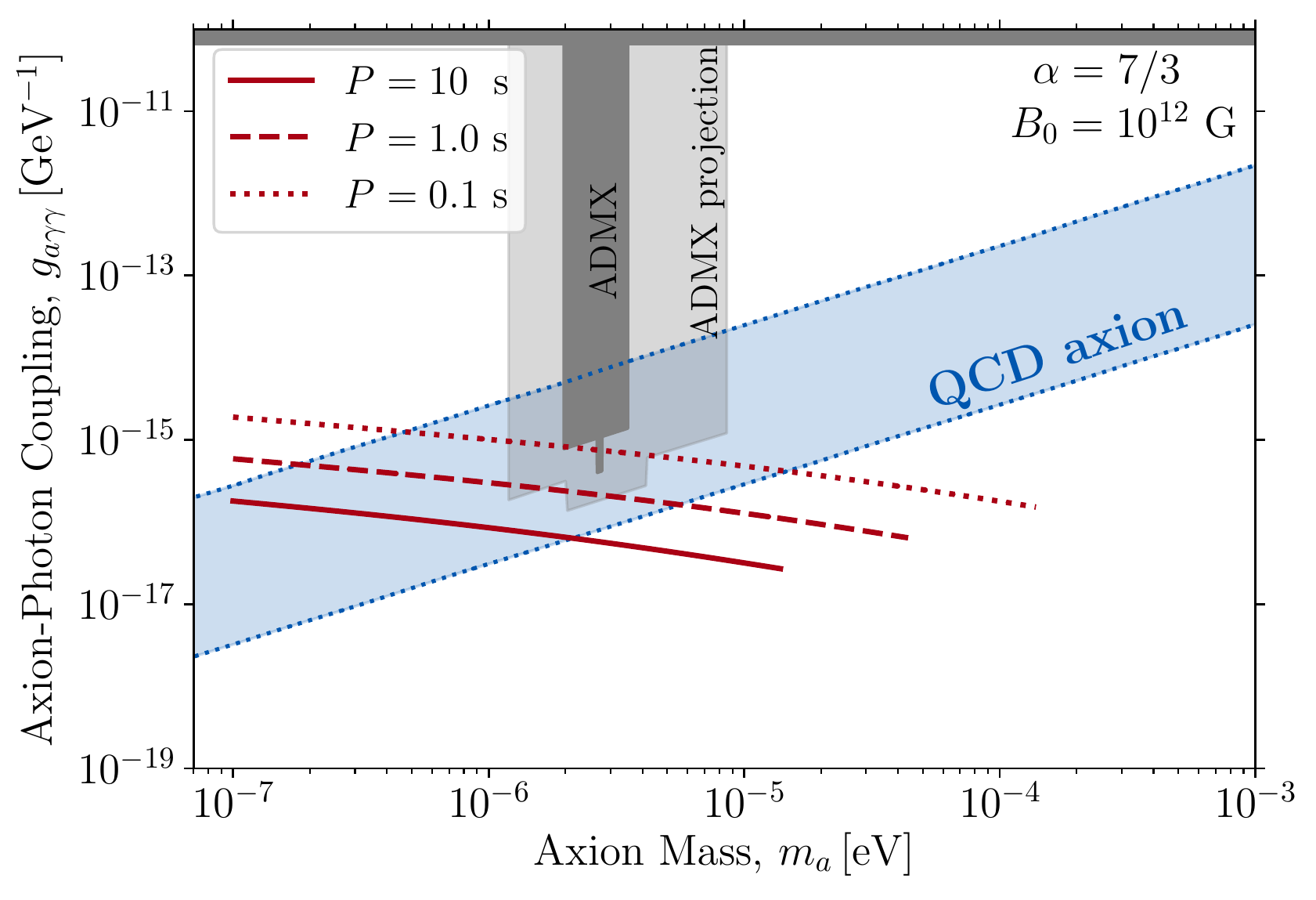}
    \caption{\textbf{Projected reach in axion-photon coupling from radio observations.} Sensitivity curves of SKA2 telescope (100 hours of observation) to the axion-photon coupling as a function of the axion mass for three values of the NS magnetic field strength (left panel) and three values of the NS spin period (right panel). Here, we assume $\alpha = 7/3$ for the slope of the DM spike, $r=3\times 10^{-9}$~pc, $d=0.01$~Gpc, and $\theta=90^\circ$.}
    \label{fig:NS_properties}
\end{figure}

In Fig.~\ref{fig:NS_properties} we report the projected sensitivity curves of SKA2 for three different values for the NS magnetic field strength (left panel) and three different values for the NS spin period (right panel), while fixing $r=3\times10^{-9}$~pc, $d=0.01$~Gpc, $\theta = 90^\circ$ and $\alpha = 7/3$. As expected, the larger the magnetic field and the spin period, the smaller the axion-photon coupling that can be probed by SKA2. Moreover, larger axion masses can be explored for larger magnetic fields or smaller spin periods. Increasing $B_0$, or decreasing $P$, causes the axion-photon conversion to occur at a larger radius (see Eq.~\eqref{eq:r_c}). The requirement that the the conversion radius is larger than the size of the NS ($r_c \geq r_{\rm NS} = 10~{\rm km}$) is then satisfied for a wider range of axion masses. 

We note that the values for $B_0$ and $P$ considered here cover almost all the possible properties of active NSs, according to the ATNF pulsar catalog~\cite{Manchester:2004bp}. On the other hand, old dead NSs are expected to have low magnetic fields and large spin periods, providing weaker sensitivities. However, their properties are quite uncertain and model-dependent~\cite{Safdi:2018oeu,FaucherGiguere:2005ny,Popov:2009jn}. In all cases, we have also verified that the plasma remains bound to the NS. Even down to the innermost stable circular orbit, the forces from the NS magnetosphere dominate over the gravitational force from the BH by many orders of magnitude.

Finally, we conclude this appendix by discussing how the radio signal depends on the model for the NS magnetosphere. In deriving the expression for the conversion radius~\eqref{eq:r_c}, we have assumed that the NS plasma is dominated by electron and positrons. The presence of ions would reduce the conversion radius once all the other quantities are fixed, since the plasma frequency~\eqref{eq:plasma_frequency} would be suppressed by the larger mass $m_c$. This would increase the radio signal according to Eq.~\eqref{eq:power} at the expense of probing smaller axion masses. Moreover, different analytic models and numerical simulations of the NS magnetosphere lead to different profiles for the plasma frequency~\cite{Petri:2016tqe}. This might increase or reduce the amplitude of the radio signal. For example, Ref.~\cite{Safdi:2018oeu} pointed out that the electrosphere model instead of the Goldreich-Julian model for the NS plasma~\cite{10.1093/mnras/213.1.43P} provides larger or smaller radio signals depending on the polar angle and the misalignment between the magnetic dipole axis and the rotation axis.

\end{document}